\documentstyle[12pt,world_sci,epsf]{article}
\def\p{{\scriptscriptstyle +}}
\def\m{{\scriptscriptstyle -}}

\begin{document}

\title{{\bf STRING MODELS FOR LOCALLY SUPERSYMMETRIC GRAND UNIFICATION}\thanks{
Talk presented by S.-w. Chung at DPF 94, Albuquerque, New Mexico}}
\author{S. CHAUDHURI\thanks{Email: sc@nsfitp.itp.ucsb.edu}\\
{\em Dept. of Physics and Institute for Theoretical Physics\\
University of California, Santa Barbara, CA 93106}\\
\vspace{0.2cm}
S.-W. CHUNG,\thanks{Email: chung@fnth04.fnal.gov} G. HOCKNEY\thanks{Email:
hockney@fnth16.fnal.gov} and J. LYKKEN\thanks{Email: lykken@fnalv.fnal.gov}\\
{\em Fermi National Accelerator Laboratory\\
P.O. Box 500, Batavia, IL 60510}}
\maketitle
\setlength{\baselineskip}{2.6ex}

\begin{center}
\parbox{13.0cm}
{ \begin{center} ABSTRACT \end{center}
{\small \hspace*{0.3cm}
Phenomenologically viable string vacua may require incorporating
Kac-Moody algebras at level $\geq 2$.
We exploit the free fermionic formulation
to construct N=(0,2) world-sheet supersymmetric string models with specific
phenomenological input: N=1 spacetime supersymmetry, three generations of
chiral fermions in gauge groups $SO(10)$ or $SU(5)$, adjoint Higgses,
and a single Yukawa coupling of a fundamental Higgs to the third
generation.  In this talk, we will show models of gauge group $SO(10)$ and
of $SU(5)$ without any gauge singlet moduli, and show some novel features
appearing in the connection of these two models.
The accompanying, and rather
non-trivial, discrete chiral sub-algebras can determine hierarchies in the
fermion mass matrix. Our approach
to string phenomenology opens up the possibility of {\it concrete}
explorations of a wide range of proposals both for dynamical supersymmetry
breaking and for the dynamics of the dilaton and other stringy moduli.
  } }  \end{center}

Model building in string perturbation is not fully explored, partly because
there exists millions of vacua (and each vacuum, if ``realistic'', is
cumbersome to analyze in full details), and partly because we have not really
tried hard enough to explore this area in an intelligent way.  For instance,
there are only three basic examples of superstring GUT constructions in the
literature\cite{dhl,fer,CCL}
in spite of thousand papers on traditional GUT models.

We have chosen to base our exploration of string vacua
on four  dimensional closed free fermionic string models,
which are heterotic superstring vacua described by a world--sheet
lagrangian for $64$ real (Majorana-Weyl) free fermions, together with two
bosons representing
the two transverse coordinates of 4-d spacetime in the light-cone formalism.
This construction is described
in detail in refs \cite{klt,abk,ab,klst}; we will, for the most part,
follow the notation and conventions of refs \cite{klst,dhl}.

Models are conveniently specified by their one-loop partition
functions, which include all the spacetime particle spectrum; these
involve a sum over spin structures:
\begin{equation}
Z_{\rm fermion} = \sum_{\alpha,\beta}\;{C^{\alpha}_{\beta}}\,
Z^{\alpha}_{\beta} \quad ,
\label{Gene}
\end{equation}
where the $C^{\alpha}_{\beta}$'s are numerical coefficients, while
$\alpha$ and $\beta$
are $64$-dimensional vectors labeling different choices of
boundary conditions for the fermions around the two independent
cycles of the worldsheet torus.
For each real fermion there are two possible choices of boundary
conditions around a given cycle: either periodic (Ramond) or
antiperiodic (Neveu-Schwarz). However for fixed $\alpha$ and $\beta$
the real fermions always pair up into either Majorana or
Weyl fermions; if a particular Weyl pairing occurs consistently
across {\it all} $\alpha$ and $\beta$, then this pair can be
regarded as a single {\it complex} fermion.
    For such complex fermions more
general boundary conditions -- any rational ``twists'' --
are then allowed\cite{klt,ab,ckt}. A useful notation denotes a
complex Ramond fermion as a $-1/2$ twist, while a general
$m/n$ twist indicates the complex fermion boundary condition
\begin{equation}
\psi \rightarrow {\rm exp}\left[2\pi i\,{m\over n}\right]\;\psi\quad .
\end{equation}

The contribution of any sector $\alpha$ to the partition
function contains a generalized GSO projection operator.
Up to an overall constant, this is given by:
\begin{equation}
\sum_{\beta}\;C^{\alpha}_{\beta}\;{\rm exp}
\left[-2\pi i\beta \cdot \hat{N}(\alpha )
\right]\quad ,
\label{GenGSO}
\end{equation}
where $\hat{N}(\alpha )$ is the fermion number operator defined in
the sector $\alpha$. There are subtleties in the proper
definition of $\hat{N}(\alpha )$ for real Ramond fermions; these
are discussed in ref.\ [7].
Thus building a fermionic string model
amounts to choosing
an appropriate set of $\alpha$'s, $\beta$'s, and $C^{\alpha}_{\beta}$'s,
then performing the GSO projections to find the physical spectrum.
These choices are greatly constrained by the requirement of
modular invariance of the one-loop partition function; in addition,
higher loop modular invariance imposes a factorization condition
on the $C^{\alpha}_{\beta}$'s.
Together these requirements imply that the $\{\beta\}$
are the same set of vectors as the $\{\alpha\}$, and that, if two
sectors $\alpha_1$ and $\alpha_2$ appear in the partition function,
then the sector $\alpha_1 + \alpha_2$ must also appear. These facts
allow one to specify the full set of $\alpha$'s and $\beta$'s by
a list of ``basis vectors'', denoted $V_i$, and we will later express
$\alpha$ in terms of these basis vectors with coefficients $\alpha_i$.
The $C^\alpha_\beta$'s parametrizing the generalized GSO projection operators
 in eq.\ (\ref{GenGSO}) can also be
reexpressed in the same basis in terms of new parameters $k_{ij}$;  these are
discussed in [4].

Models that consist of real fermions can have
higher level Kac--Moody algebra and gauge group of rank lower
than 22, which is an advantage if one wants to build up a realistic
model.  However, the spacetime particle spectrum in such models are
generally realized in rather intricate fashion, and quite different
from level one models.  For example, the 45 gauge
bosons of $SO(10)$ at
level 2 in the model we will later present reside in both
the untwisted sector and 7 other twisted sectors.  Furthermore,
there are probably 200 twisted sectors containing massless spectrum
before GSO projection in a typical model of three generations of chiral
fermions ${\bf 16}_L$.
   Fortunately, these complications and tedious checks can be
easily handled within  seconds by our newly developed symbolic manipulation
computer package.  The details  regarding this package will be
explained in our later publication\cite{cchl}.

Equipped with this powerful tool, one can start to explore the string
vacua which incorporate a considerable  amount of low--energy
phenomenology input.  Studies along this direction can help us to extract some
interesting features about true string vacua which are closer
to nature, and it may also reveal hints for eventual
non-perturbative formulation of string theory.
To give one example besides models with three generations of chiral fermions
and adjoint scalars\cite{CCL}, we will present
a model with N=1 spacetime SUSY, three generations of chiral
fermions in gauge group $SO(10)$ at level two, and {\it no moduli}
(except dilaton).  Moreover, we will show, from this $SO(10)$ model, one can
obtain a model of $SU(5)$ at level two with three chiral 10s
by just tuning parameters $k_{ij}$ in the
GSO projection operators.

The first model
has  basis vectors $V_0$ -- $V_9$ specified as follows:
\begin{eqnarray}
V_0&=(11111111111111111111\|
111111111111\vert 11111111111111\vert 111111111111111111)
\nonumber \\
V_1&=(11100100100100100100\|
000000000000\vert 00000000000000\vert 000000000000000000)
\nonumber \\
V_2&=(00000000000000000000\|
111111110000\vert 11111111000000\vert 000000000000000000)
\nonumber \\
V_3&=(00000000000000000000\|
000000000000\vert 00001111111100\vert 000000000000000000)
\nonumber \\
V_4&=(00000000000000000000\|
110000111111\vert 11001100110011\vert 000000000000000000)
\nonumber \\
V_5&=(11100100010010010010\|
111100001100\vert 10101010101010\vert 111000000000000000)
\nonumber \\
V_6&=(11010010100100001001\|
111100001100\vert 10100101101001\vert 000001110000000000)
\nonumber \\
V_7&=(11001001001001100100\|
111100001100\vert 11110000111100\vert 000000001100000000)
\nonumber \\
V_8&=(00101101110110 0\p \m 0\p \m\|
00000000\p \p \p \p\vert 010101010101 01\vert 00 011 1\p\p \p\p\p\p\m\m0000)
\nonumber \\
V_9&=(00 \p\m 0 \p\m 0 \p\p 1 \p\p 1 000 000\|
000000000000\vert 0000 1111 0000 11 \vert 0011001100\p\p\p\p \p\p\p\p )
\nonumber
\end{eqnarray}

``1'' in the above vectors denote the value ``$-1/2$'', and ``$\pm$'' denote
the values of ``$\pm1/4$''.
The first $20$ components up to the double vertical lines denote the boundary
conditions for the right-moving fermions.
The first pair of right-movers are spacetime fermions
(corresponding to the superpartners of the two transverse
directions in 4-d), while
the other $18$ right-movers are ``internal''.  The remaining $44$
components are left-movers.  The first $12$ components on the right of
the double vertical lines denote the quantum numbers under
$SO(10) \times U(1)$, and the
second $14$ components are real fermions necessary for this particular
$SO(10)$ embedding.

In fermionic string models, there exists an ``untwisted sector'',
with $64$ Neveu-Schwarz Weyl fermions. The untwisted sector
contains the graviton, dilaton, antisymmetric tensor field, and some of the
gauge bosons.  The requirement
of a worldsheet supercurrent constructed out of the right-movers
and the spacetime bosons is a consistency constraint on model
building. As a result, $V_1$ contains massless gravitinos, corresponding to
an $N$$=$$4$ spacetime supersymmetry before the GSO projection. After the
projection one only has $N$$=$$1$ spacetime SUSY.  Superpartners of states
in some sector $\alpha$ will be found in $V_1+\alpha$.
$V_0$ is required in all fermionic string models by modular invariance.
To produce a model with $SO(10)\times U(1)$ and three generations of chiral
fermions, we also have to choose $k_{ij}$ with $i> j>0$ or $i=j=0$,
\begin{equation}
k_{00}=1/2,~~k_{61}=1/2,~~k_{73}=1/2,~~k_{83}=1/2,~~k_{86}=1/2,~~
     {\rm and~the~rest}=0.
\end{equation}

The $45$ gauge bosons of $SO(10)$ at level 2 reside in the untwisted sector,
$V_2$, $V_3$, $V_4$, $V_2+V_3$, $V_2+V_4$, $V_3+V_4$, and $V_2+V_3+V_4$.
One can check that the root vectors forming from the ``fermionic charges''
associated with the first 12 left-moving fermions, or first six complex
fermions, do have length 1; thus it is a level 2 Kac--Moody algebra.
Three generations of $16_L$ are contained in $\{ V_5, V_6, V_7\} +
\{ 0, V_2, V_4, V_2+V_3, V_2+V_4, V_3+V_4, V_2+V_3+V_4 \}$.  Notice that
$V_{5,6,7}+V_3$ are projected out.  In this model, the observable gauge group
is $SO(10)\times U(1)^4$, and the hidden gauge group is $U(1)^3\times U(2)$.
There are no moduli in this model, {\it i.e.}, no gauge singlet with
respect to both observable and hidden sectors.  However, we do have 8 states,
half of them  contained in $V_3$ and the other half in $V_3+2*V_9$,
which are gauge singlets with respect to the observable gauge group but not
gauge singlets under the hidden gauge group.
It is worthwhile to point out
that in this construction one uses $26$ fermions with central charge $c=13$
to construct a conformal field theory based on Kac--Moody algebra of
$SO(10)$ at level 2 tensoring $U(1)$,
whose total central charge is $c=10$.  The precise form of the remaining
chiral algebra could be worked out.

Now if one chooses the same set of $k_{ij}$ values except changing
$k_{93}=1/2$, one would obtain a model with observable gauge group
$SU(5)\times U(1)^5$, while the hidden gauge group is unchanged.  The
gauge bosons associated with $V_3, V_2+V_3, V_3+V_4, V_2+V_3+V_4$ do not
pass the GSO projection.  The remaining gauge bosons form exactly  the
root lattice of $SU(5)$, and the level is still $2$ because the root
length is unchanged.  It is interesting to point out that the original
${\rm 16}_L$'s of $SO(10)$ from $V_{5,6}$ and their gauge partners
become ${\rm 10}_L+{\bar{\rm 5}}_L+{\rm 1}_L$ of $SU(5)$, while the
${\rm 16}_L$ from $V_{7}$  become ${\rm 10}_L+{\bar{\rm 5}_R}+{\rm 1}_R$,
and an additional ${\bar{\rm 5}}_L+{\rm 1}_L$ appears in
$ V_7+V_9  +\{V_2+V_3,V_3+V_4, V_2+V_3+V_4 \} $.

Our experience in model building tells us that it is not difficult to
get models with three generations of chiral fermions with adjoint
Higgs, or with no moduli, or with only one Yukawa coupling to the
third generation.  But it may take some effort to construct a model with
all the desired features.

\def\cmp#1{{\it Comm. Math. Phys.} {\bf #1}}
\def\pl#1{{\it Phys. Lett.} {\bf #1B}}
\def\prl#1{{\it Phys. Rev. Lett.} {\bf #1}}
\def\prd#1{{\it Phys. Rev.} {\bf D#1}}
\def\prr#1{{\it Phys. Rev.} {\bf #1}}
\def\prb#1{{\it Phys. Rev.} {\bf B#1}}
\def\np#1{{\it Nucl. Phys.} {\bf B#1}}
\def\ncim#1{{\it Nuovo Cimento} {\bf #1}}
\def\jmp#1{{\it J. Math. Phys.} {\bf #1}}
\def\mpl#1{{\it Mod. Phys. Lett.} {\bf A#1}}

\bibliographystyle{unsrt}

\end{document}